\documentclass[twocolumn,letter]{revtex4}
\usepackage{depken}
\usepackage{graphicx}
\usepackage{psfrag}

\newcommand{\note}[1]{{#1}}
\newcommand{\tr}{{\rm tr}}
\bibstyle{apsrev}
\begin{document}
\author{Martin Depken}
\affiliation{Instituut-Lorentz for Theoretical Physics, Leiden University\\
P.O. Box 9506, NL-2300 RA Leiden, The Netherlands}
\author{Martin van Hecke}\affiliation{Kamerlingh Onnes Lab, Leiden
University, PO box 9504, 2300 RA Leiden, Netherlands}
\author{Wim van Saarloos}
\affiliation{Instituut-Lorentz for Theoretical Physics, Leiden University\\
P.O. Box 9506, NL-2300 RA Leiden, The Netherlands}
\title{Continuum approach to wide shear zones in quasi-static granular matter}

\begin{abstract}
Slow and dense granular flows often exhibit narrow shear bands,
making them ill-suited for a continuum description. However, smooth
granular flows have been shown to occur in specific geometries such
as linear shear in the absence of gravity, slow inclined plane flows
and, recently, flows in split-bottom Couette geometries. The wide
shear regions in these systems should be amenable to a continuum
description, and the theoretical challenge lies in finding
constitutive relations between the internal stresses and the flow
field. We propose a set of testable constitutive assumptions,
including rate-independence, and investigate the additional
restrictions on the constitutive relations imposed by the flow
geometries. The wide shear layers in the highly symmetric linear
shear and inclined plane flows are consistent with the simple
constitutive assumption that, in analogy with solid friction, the
effective-friction coefficient (ratio between shear and normal
stresses) is a constant.  However, this standard picture of granular
flows is shown to be inconsistent with flows in the less symmetric
split-bottom geometry --- here the effective friction coefficient
must vary throughout the shear zone, or else the shear zone
localizes. We suggest that a subtle dependence of the
effective-friction coefficient on the orientation of the sliding
layers with respect to the bulk force is crucial for the
understanding of slow granular flows.

\end{abstract}
\maketitle

\section{Introduction}
Granular flows show a very wide variety of behaviors, and while the
microscopic dynamics of dry cohesion-less grains is simple and well
understood, there is no general theory describing their emergent
macroscopic properties. Flowing grains can roughly be classified
into three regimes by the relative importance of inertial
effects~\cite{Jackson86}. For strong external driving the grains
form a gaseous state. Here particle interactions are dominated by
binary collisions, and this regime is well captured by modified
kinetic theories~\cite{Jenkins83,Haft83,Campbell90,Goldhirsch99}. On
lowering the driving strength the flow becomes denser, with
collisions becoming correlated and often involving several particles
at once. In this regime inertia is still important, but kinetic
theories become increasingly difficult to justify and
apply~\cite{Bizon99}. On further lowering the driving strength, the
granular media enter the quasi-static regime where inertial effects
are negligible. The grains form enduring contacts, leading to highly
complex contact and force networks. The modeling of these flows is
still in its infancy, and there is no {\em general} approach which,
for given geometry and grain properties, predicts the ensuing flow
fields.

Most slow granular flows can not be considered smooth. Their flow
fields vary strongly on the grain scale. For example, in many
experimental realizations one observes shear localization where the
flow of the material is concentrated into a very narrow shear
band~\cite{Nedderman92,Bocquet01,Veje99,Douady99}. In such
situations the flow can be modeled as two solid blocks sliding past
each other. If the shear stress, $\tau$, is simply proportional to
the normal pressure, $P$, then these materials are referred to as
ideal cohesion-less Coulomb materials. Many formulations of granular
flow focus immediately on this narrow shear band regime
\cite{Douady99,Bocquet01}.

\begin{figure}[htp]
\includegraphics[width=\columnwidth]{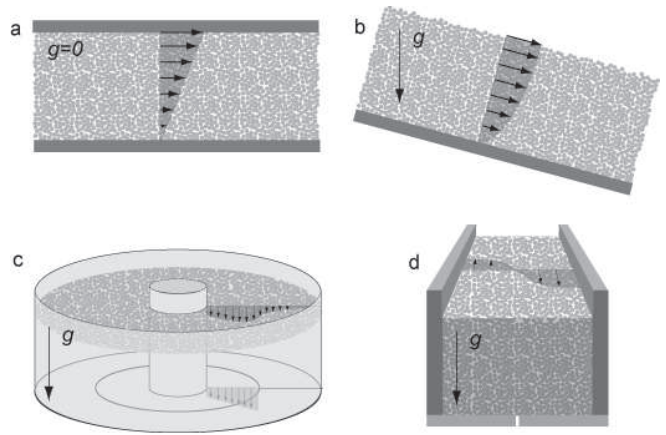}
\caption{\label{fig:setups}Geometries in which smooth, quasi-static
grain flows occur. The velocity field is sketched with arrows. (a)
Linear shear in absence of gravity. (b) Inclined plane flow close to
the critical inclination angle. (c) Taylor-Couette flow with split
bottom.  (d) Linear shear over a split bottom.}
\end{figure}

However, there are a number of systems that display smooth velocity
fields and wide shear zones. These should be amenable to a continuum
description. Among these is the planar-shear cell without gravity,
which has been examined numerically
in~\cite{Midi04,Babic90,Thompson91,Zhang92,Schwarz98,Aharonov02}
(Fig.~\ref{fig:setups} a). Another example is slow flows down
inclined planes, which in simulations of 3D systems appear to reach
a quasi-static state~\cite{Silbert01} (Fig.~\ref{fig:setups} b).
Though conceptually simple, both of these situations are hard to
realize experimentally. In recent experiments, using a modified
Taylor-Couette cell with split bottom (Fig.~\ref{fig:setups} c),
robust and wide shear zones were obtained in the quasi-static, dense
regime \cite{Fenistein03,Fenistein04,Fenistein05}. While we do not
expect there to be a "universal" continuum theory of granular flow,
these observations strongly suggests that there is a continuum
theory with its own domain of validity, that should capture this
smooth quasi-static granular flow regime.

Our approach is to test whether a straight-forward continuum model
of these smooth flow fields, based on a minimum of readily testable
physical assumptions, can be made consistent with the numerical and
experimental data available for smooth quasi-static flows. In
addition to mass, linear- and angular-momentum conservation, we need
to find additional relations between the six components of the
stress tensor $\sigma_{ij}$ and the state of the system
characterized by such quantities as strain history, packing fraction
etc. Such constitutive equations are particularly simple for
Newtonian fluids (leading to the Navier-Stokes equation) and elastic
solids (leading to the equations of linear elasticity). The yielding
behavior of granular media illustrates that the constitutive
equations must here take on a more complicated form. Granular media
are athermal and dissipative --- hence, when no external energy is
supplied, grains {\em jam} into a rigid, solid-like state which can
sustain a finite load before yielding \cite{Liu01}. Grains are made
to flow by supplying an external (shear) stress to overcome this
yielding threshold. As a result, for very slow and dense granular
flows, the shear stresses are finite and do not approach zero.  This
complicates matters considerably.

Our approach has two important ingredients, the details of which can
be found in Section \ref{sec:qs}. Firstly, we are guided by the well
known fact that dense grain flows exhibit rate
independence~\cite{Midi04}. For the velocity fields this means that,
to good approximation, the entire velocity profile scales
identically with the external driving: When the driving speed is
doubled, the whole velocity field doubles. The stresses are also
approximately rate independent, meaning that when the speed is
doubled, the stresses stay the same. This makes the relation between
the stresses and the flow field rather special, and even if we could
determine the full stress field, we could never hope to get the full
velocity profile. The approach we take relates the stresses to
certain aspects of the {\em geometry} of the flow. This results in
statements regarding material sheets in the flow, within which the
particles on average only perform a collective rigid body motion
with respect to each other. A trivial instance of such sheets are
the layers of constant velocity present in the linear setups in
Fig.~\ref{fig:setups} a, and b, and illustrated in
Fig.~\ref{fig:plane}. In order to say something about the actual
velocities of such planes one would have to appeal to a sub-dominant
dependence on shear rate.

Secondly, when the grains are flowing, they experience large
fluctuations~\cite{Komatsu00}. Hence, we assume that if in a certain
plane the strain rate is zero, then there will be no residual shear
stress in this plane --- if there was a shear stress, there would be
a shear flow. Hence, all shear stresses are dynamically sustained,
and there are no elastic shear stresses. Thus, we will not attempt
to model a mixture of solid and flowing behavior as done
in~\cite{Volfson03}. This implies that the principal- strain and
stress directions are the same (see Section~\ref{sec:relax}).

In Section~\ref{sec:sz} we apply this framework to the four
geometries depicted if Fig.~\ref{fig:setups}. In the linear
geometries (Fig.~\ref{fig:setups} a, and b), symmetry considerations
directly give that the principal directions are constant throughout
the system, and thus the equations are automatically closed. This
gives the standard Mohr-Coulomb relation $\tau=\mu P$, with the
effective-friction coefficient necessarily constant throughout the
sample. For the less symmetric geometries (Fig.~\ref{fig:setups}, c
and d) the local orientation of the above material sheets will vary
throughout the cell. This allows us to separate the effect of
constitutive assumptions regarding the rates in the system and the
geometry of the flow (see Section~\ref{sec:SFS} for further
details). If we maintain that the effective friction coefficient is
constant throughout the sample, we find that the shear zones have
infinitesimal width.  {\em The standard approach of a constant
effective-friction coefficient between shearing planes fails} (see
Section~\ref{sec:curved}, and especially Fig.~\ref{fig:pict}). In
fact, we then completely recover the prediction regarding the
location of the shear zone that was derived on the basis of torque
minimization by Unger {\it et al.} in~\cite{Unger04}.

To capture the experimentally observed widening of the shear zone in
the bulk, the effective-friction coefficient has to vary throughout
the shear zone. \note{We argue that this can only be done through a
dependence of the effective-friction coefficient on the orientation
of the shearing surface with respect to any bulk force (here
gravity). The possible origin of such an angle dependence is
discussed in Section~\ref{sec:discuss}.}

 \section{quasi-static granular flows}
\label{sec:qs} At the heart of the present development lies the
peculiar fact that in order for a granular matter to support a
shearing state, no matter how slow, a finite shear stress is needed.
This is reminiscent of solid friction, but in stark contrast with
the situation in Newtonian fluids. This feature is clearly visible
in the experimental results reported
in~\cite{Hartley03,Midi04,Aharonov99,Miller96}.

\subsection{Explicit rate independence}
The strain-rate tensor $\vv D=(\v \nabla \v v+(\v \nabla \v
v)^\dagger)/2$ plays a central role in ordinary fluid
mechanics~\cite{Astarita74}. The use of only the symmetric part of
the deformation-rate tensor $\v \nabla \v v$ ensures that no
stresses are induced by pure local rotations of the material
(principle of material objectivity). In the theory of simple fluids
one assumes that the knowledge of the complete history of $\vv D$,
for any material point, will give the stresses at that point.
\note{In the case that there are no memory effects this means that
the stress tensor can be expressed as an isotropic tensor function
of the strain rate tensor. For such functions, the first
representation theorem (Rivlin-Ericksen theorem~\cite{Philippe88})
states that the most general constitutive equation can be written as
\be{eq:rept}
  \vv \sigma=\alpha_0 \vv I +\alpha_1 \vv D+\alpha_2\vv D^2
\ee with $\alpha_i=\alpha_i(I_{D},II_{D}, III_{D})$, and the
invariants
$$
  I_{D}=\tr \vv D, \quad II_{D}=\tr \vv D^2,
  \quad III_{D}=\det \vv D.
$$
As mentioned above, the granular flows that we want to describe are
such that we have finite shear stresses even as the shear rate
approaches zero. Thus, we can split the stress tensor, $\vv\sigma$,
into a rate-independent part, $\vv\sigma_0$, and a rate-dependent
part, $\vv\sigma_1$,
\be{eq:str}
   \vv\sigma=\vv\sigma_0+\vv\sigma_1,
\ee
in such a way that $\vv\sigma_0$ is {\em not} proportional to the
identity operator (i.e. it contains shear stresses, and is hence not
simply a hydrostatic pressure), and $\vv \sigma_1$ vanishes as the
strain rates approaches zero. The condition on $\vv \sigma_0$
directly tells us that in the zero shear rate limit $\alpha_1$
and/or $\alpha_2$ must be singular.}

Theoretically there exist a flow regime in which $\vv\sigma_0$ alone
sets certain properties of the flow. We will refer to this regime as
quasi-static (the precise experimental definition is given in
Section~\ref{sec:dim}). The information that can be extracted from
the rate-independent part of the stress tensor will in general be of
the type specifying, e.g., constant velocity surfaces. Due to rate
independence, questions regarding the magnitude of the velocity
field can not be answered by considering this limit alone, and
neither can questions regarding the stability of any wide shear
zones.

We further assumes that there are only local interactions in the
bulk (principle of local action)~\footnote{Non-local interactions
would be possible if a macroscopic length scale is present in the
bulk. This could happen for a system close to something reminiscent
of a second-order phase transition. The existence of such a
transition point, especially in the context of ``jamming'', where
the mentioned critical point is often referred to as point $J$, is
at present under active
investigation~\cite{Liu98,Silbert05,OHern03,Liu01}. The location of
point $J$ would be precisely at the packing fraction where the
material starts to support an external shear load. The systems we
hope to describe supports finite shear stresses even as the shear
rate approaches zero, and we are hence away from point $J$.}. The
implicit assumption in our approach is that our continuum
description is valid, upon coarse graining over some small but
finite length scale.

As for the history dependence, the systems we consider are such that
the flow direction is also a symmetry direction. For such system in
steady shearing states, any material point will always have the same
surrounding flow field. Thus $\vv D$ does not change \note{(up to a
rotation)} along the evolution paths of the material elements, and
memory effects are washed out.

The strain-rate tensor is symmetric, and is hence completely
specified by six parameters. We can choose these parameters as,
e.g., the principal strain rates, and the orientation of the
principal directions (specified by three angles). Using this
parameterizations of the strain-rate tensor enables us to isolate
the rate dependence from the orientational dependence. We denote the
principal strain rates with $\gamma_i$, and the three angles
defining the principal directions by $\theta_i$. The angles are to
be taken with respect to some suitably chosen local reference
direction (e.g. the gravitational field). Then, the general form of
the stress tensor is
\be{1}
  \vv\sigma_0=\vv\sigma_0(\gamma_1,\gamma_2,\gamma_3,\theta_1,\theta_2,\theta_3,\ldots),
\ee where the dots indicate a possible dependence on parameters not
directly related to the shear. Rate independence of the stress
tensor implies invariance under the re-scaling $\v v\rightarrow b\v
v$, and consequently invariance under $\vv D \rightarrow b\vv D$
($\Leftrightarrow\,\, \gamma_i\rightarrow b \gamma_i$). Therefore
the stresses can only depend on the ratios of the principal-strain
rates, and not the strain rates themselves. Hence, \be{eq:rindep}
  \vv\sigma_0=\vv\sigma_0(\gamma_1/\gamma_3,\gamma_2/\gamma_3,\theta_1,\theta_2, \theta_3,\ldots).
\ee
To proceed with a general theory one would need to include the full
dependence on the principal-strain-rate  ratios. The flows we will
consider are of a limited type, which enables us to study the
influence of the angles $\theta_i$ without specifying the dependence
on the principal-strain-rate ratios. We now proceed by clarifying
his point.

\subsection{Shear-free sheets}
\label{sec:SFS} The systems we wish to consider are all such that,
on the scale of the coarse graining, one can think of them as
consisting of  material sheets, with no internal shear, shearing
past each other. We will here make this more precise and derive some
important consequences. In Section~\ref{sec:dim} we argue that the
density in any flowing region is essentially constant, and thus mass
conservation ensures that the flow is divergence free. This will be
assumed already here.

\begin{figure}
  \includegraphics[width=\columnwidth]{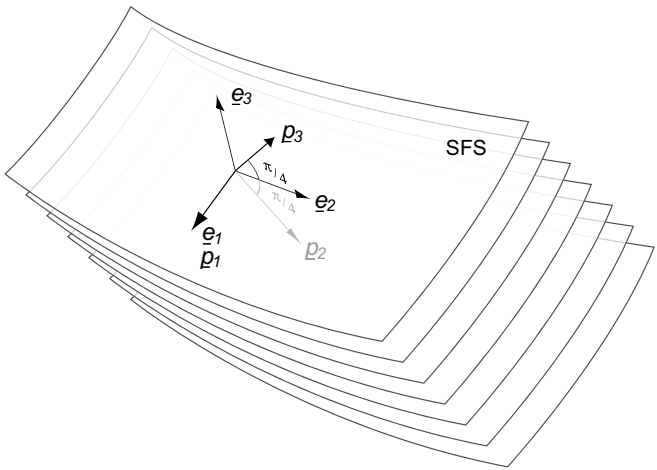}
  \caption{\label{fig:SFS} Pictorial view of the partitioning of space
  into a set of SFS. Also illustrated is a specific instance of the SFS basis,
  and the corresponding principal-strain basis.}
\end{figure}

Consider a system for which it is possible to find a reference frame
such that the velocity field is time independent (e.g. in the center
of mass frame). We define a flow sheet as a surface in the flow,
such that if a material point starts out on the surface, it stays on
the surface throughout the time evolution of the system. If there
are no strains within the sheet, we will refer to it as shear-free
sheet (SFS). That is to say that the restriction of the strain-rate
tensor to the sheet vanishes. The flows treated later are such that
the whole shearing region can be divided into a collection of SFS
(the SFS form a foliation of space occupied by the shear band). In
any orthogonal and normalized basis field, with the two first basis
vectors tangential to the SFS, the component form of the strain
tensor is
\be{eq:Dk}
 (\vv D)=\left(\begin{array}{ccc}0 & 0& d_1\\
0& 0& d_2\\ d_1&d_2&0
\end{array}\right).
\ee
Here we have used the fact that the total flow is assumed to be
divergence free, $\v \nabla\cdot\v v={\rm tr}\vv D=0$. Hence, the
principal-strain rates are $\gamma_1=0$ and $\gamma_{2,3}=\pm
\sqrt{d_1^2+d_2^2}$. The major advantage of considering these flows
is now obvious: The ratio between the principal strain rates remain
constant throughout the system (even though $d_1$ and $d_2$ are free
to vary). Thus we can drop this dependence in stress tensor,
giving
\be{}
  \vv\sigma_0=\vv\sigma_0(\theta_i,\ldots).
\ee
The strain-rate ratios will be dropped from now on, and the ability
to do so is crucial for the rest of the development. This enables us
to probe the angle dependence alone. By a simple rotation in the
SFS, the component form of the strain-rate tensor can be recast as
\be{eq:De}
  (\vv D)_{\rm SFS}=\left(\begin{array}{ccc}0 & 0& 0\\ 0& 0& \dot\gamma\\
  0&\dot\gamma&0 \end{array}\right),\quad \dot\gamma=\sqrt{d_1^2+d_2^2}.
\ee
We will refer to the basis that realizes this component form of the
strain-rate tensor as the SFS basis, $\{\v e_1, \v e_2,\v e_3\}$.
\note{Viscometric flows~\cite{Astarita74} have this form of the
strain rate tensor, but since we will put much emphasis on the
physical picture offered by the SFS, we will continue to refer to
these flows as SFS flows.} This simple form tells us that the shear
between planes is always directed along $\v e_2$. Hence, for these
flows we have a picture consistent with the SFS sliding past each
other (see Fig.~\ref{fig:SFS}). For later reference the
principal-strain basis (the basis spanned by the eigenvectors of
$\vv D$) $\{\v p_1, \v p_2, \v p_3\}$ is easily seen to be given as
\be{eq:bsb}
  \v p_1=\v e_1, \quad \v p_{2,3}=(\v e_2\mp \v e_3)/\sqrt{2},
\ee
in terms of the SFS basis (see Fig.~\ref{fig:SFS}).
We now proceed to argue for a specific form of the stress tensor in
these two bases.
\subsection{Stress relaxation}
\label{sec:relax} We claim that in the principal-strain basis the
stress tensor takes the form
\be{eq:stress0}
  (\vv\sigma_0)_{{\rm P}}=\left(\begin{array}{ccc}P^1&0&0\\0&P^2&0\\0&0&P^3
  \end{array}\right), \quad P^i=P^i(\theta_j,\ldots).
\ee
To justify this we argue that force fluctuations are rapid in
shearing flows. At any instance, two neighboring fluid elements,
positioned relative to each other along any of the principal-strain
directions, perform only a collective rigid-body, and a relative
stretching movement. Since these material points are not shearing,
no shear forces should be generated between them. If any such forces
are present as the material points enter this no-shear
configuration, we assume them to relax fast enough to be ignored.
The assumption that the principal directions of strain and stress
are aligned is also central to the flow rules for many of the
existing continuum models of granular flow~(see \cite{Jackson86} and
references therein). In the SFS basis the stress tensor takes the
form
\be{eq:stressr}
  (\vv\sigma_0)_{{\rm SFS}}=\left(\begin{array}{ccc}P' &0&0\\0&P&\tau\\0&\tau&P\end{array}\right),\quad \left\{\begin{array}{l}P'=P^1\\ P=\frac{1}{2}(P^2+P^3)\\ \tau=\frac{1}{2}(P^2-P^3),\end{array}\right.
\ee
 which again makes the connection to solid friction between the
SFS. So, the introduction of the SFS enables us to construct a
physically relevant analogy with solid friction, which, as will be
shown below, yields testable predictions.  This is crucial for what
remains.

\subsection{The continuity equation}
Mass conservation, and the fact that we assume the packing fraction
to be constant throughout the shearing region, implies that the
velocity field is divergence free.  The linear-momentum continuity
equation, in conjunction with angular-momentum continuity (and the
requirement that there is no torque body couple), ensures that the
stress tensor is symmetric, $ \vv\sigma= \vv\sigma^\dagger$. This is
something we have already implicitly assumed above. The linear
momentum continuity equation reads \be{eq:a}
  \frac{\d({\rho \v v})}{\d t}+\v \nabla \cdot \vv\Pi=\v F,
\ee
 were $\v F$ is the body force, and the momentum-flux tensor is
defined as $\vv \Pi=\rho \v v \v v+\vv\sigma$. As we are interested
in quasi-static flows, we will neglect the $\O(|\v v|^2)$ term in
the definition of the momentum-flux tensor. We will further only be
interested in steady flows, and under these conditions the
continuity equation for linear momentum takes the form of a  force
balance equation \be{eq:qscons}
  \v \nabla\cdot\vv\sigma=\v F.
\ee The number of additional equations needed to close such a system
is dependent on the symmetries present, and will be addressed for
the four geometries considered below. We now turn to dimensional
analysis to determine what the relevant dimensionless parameters
are.

\subsection{Dimensional analysis and additional assumptions}
\label{sec:dim} Obvious local parameters for the flow are the volume
fraction $\phi$, the material density $\rho_{\rm m}$, the different
local stresses in the SFS basis, the particle diameter $a$, and the
shear rate $\dot \gamma$. There is a further possibility that the
local bulk-force influences the shear stress differently depending
on how it is oriented with respect to the SFS. If this is the case
we must retain a dependence of the stress tensor on the orientation
of the principal-strain basis with respect to the bulk force. This
is encoded in the angles $\theta_i$, and since there is a one-to-one
correspondence between the principal-strain basis and the SFS basis
we can take these angle to be defined with respect to the latter
instead of the former. We can now form the dimensionless quantities
\be{eq:dimlessp}
  \phi, \quad \mu=\tau/P,\quad \nu=P'/P,\quad\theta_1,\,\theta_2,\,\theta_3,
\ee
as the shear rate is taken to zero.\note{ Since the grains are hard
we assume the packing fraction to be independent of the pressure
ratios, as well as the angles. Hence, we take the packing fraction
to be constant through the quasi-static regime, and we drop it from
the development. Experimental and numerical justifications for this
are referred to in Section~\ref{sec:discuss}.} The only
dimensionless quantity that can be constructed with the strain rate
is
\be{eq:I}
  I=\frac{\dot \gamma a}{\sqrt{P/\rho_{\rm m}}}.
\ee
It was shown in~\cite{Midi04} that $I$ is the essential parameter
determining how the material flows. Quasi-static flow is to be
expected for $I$ of order $10^{-3}$ or less.

\label{eq:second} As an aside we mention that for a general $I$ in a
SFS system (we assume $\dot\gamma>0$ for simplicity) we can write
\be{}
  \vv\sigma_1=\vv\sigma-\vv\sigma_0=\Delta P' \vv I+\Delta\tau (\vv D/\dot\gamma)+\Delta P(\vv D/\dot\gamma)^2,
\ee
where the second equality defines the coefficients $\Delta P'$,
$\Delta P$, and $\Delta \tau$ through~(\ref{eq:rept}). It is also
clear that the coefficients all must vanish for vanishing shear
rates. For a SFS system only one of the fundamental invariants is
non-zero,
\be{}
  I_D=0, \quad II_D=\dot\gamma^2,\quad III_D=0,
\ee
and we have
\be{eq:}
  \vv\sigma_1=
  \left(\begin{array}{ccc}\Delta P'(\dot\gamma)&0&0\\
  0&\Delta P(\dot\gamma)&\Delta\tau(\dot\gamma)\\
  0&\Delta\tau(\dot\gamma)&\Delta P(\dot\gamma)\end{array}\right).
\ee
Hence we conclude that the general form of the stress tensor is
preserved even for finite shear rates. These predictions should all
be possible to check by simulating these systems. The above forms
should also be useful when considering the stability of these flows,
and the velocity field on the SFS. Neither are investigated further
in the present paper. Instead we continue to focus on the SFS in the
quasi-static regime.

Returning to the rate independent case, and in analogy with solid
friction, we make the additional assumption that the in-plane
pressures of SFS do not affect the friction between the SFS. Thus we
assume $\mu$ and $\nu$ to be independent. That is, the equation of
state, relating all the dimensionless parameters, splits into two
separate equations
\be{ang_dep}
  \mu=\mu(\theta_i), \quad \nu=\nu(\theta_i),
\ee
where the actual forms depend on the material. We therefore have
four independent quantities, say $P$, $\theta_1$, $\theta_2$, and
$\theta_3$, over a three dimensional space.

Alternatively, if we had the full velocity field  of some suitable
system, say through numerical simulations, we could calculate the
principal-strain basis and check that the stress tensor has the
appropriate form~(\ref{eq:stress0}). If this turns out to be true,
we can gain information about the material functions $\mu$ and $\nu$
by comparing the stresses.
\section{Flows with wide shear zones}
\label{sec:sz}
The four systems we consider (see
Fig.~\ref{fig:setups}) all display wide shear-zones and are also
easily identified as SFS flows. We start by considering two systems
with a high degree of symmetry, and then move on to the rather
non-trivial split-bottom Couette geometries.
\subsection{Planar shear, and inclined plane geometries}
\label{sec:incp} 
\begin{figure}
\includegraphics[width=\columnwidth]{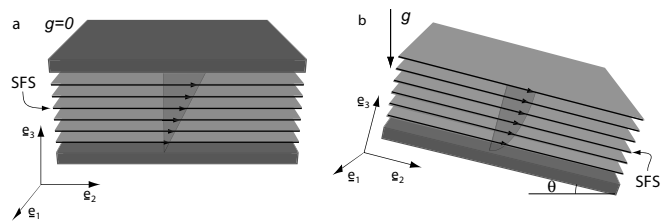}
\caption{\label{fig:plane} The plane shear cell, and the inclined
plane geometry, with the SFS as well as the SFS bases indicated.}
\end{figure}

The first two geometries, to which we apply the
above, are those of the linear shear cell without a gravitational
field, and   the inclined plane in a gravitational field (see
Fig.~\ref{fig:plane}).
Due to the symmetry present in both geometries, we can directly
identify the SFS as being parallel to the boundaries. Hence, $\v
e_3$ is always perpendicular to these. The only shear present
between the planes is in the direction of the velocity, so $\v e_2$
points in the flow direction, and $\v e_1=\v e_2\wedge\v e_3$.  The
strain-rate tensor has the expected form~(\ref{eq:De}) with
$\dot\gamma=\frac{\d v}{\d x^3}$. In the considered geometries the
principal-strain directions are constant throughout the sample, and
thus the angles $\theta_i$ are also constant.

The plane-shear geometry is trivial in that all the elements of the
SFS basis constitute a bulk-symmetry direction. Hence, all relevant
parameters must be constant throughout the bulk. Symmetry alone has
already fixed the SFS planes, and appealing to subdominant
dependence on $I$, we conclude that the shear rate is constant. This
gives a linear velocity profile under the assumption that the
boundaries do not break the symmetry by inducing localization of the
shear zone.

In the inclined-plane geometry the $\v e_3$ direction is no longer a
bulk symmetry direction, and thus $I$ is not constant along $\v
e_3$. Hence we have a more complicated velocity profile. From
equation~(\ref{eq:qscons}), or by simple force balance arguments
between the SFS, we have $\mu=\tan \theta$ (where $\theta$ is
introduced in~Fig.~\ref{fig:plane}). Since the requirement on the
effective-friction coefficient to be constant is geometrical in
origin, it holds true to all orders in $I$. Including a sub-dominant
dependence on $I$ in the effective-friction coefficient,
$\mu=\mu(\theta,I)=\tan \theta$, thus tells us that $I$ is constant
throughout the sample. This gives the well known Bagnold
profile~\cite{Silbert01,Midi04,Bagnold54}.

Further, in this system the numerical results of~\cite{Silbert01}
(the small inclination setups in three dimensions) show a linear
relation between the pressures. It is also seen that the pressures
$\sigma_{22}$ and $\sigma_{33}$ are very close to equal, while
$\sigma_{11}$ differs substantially from the others. In the above
treatment we have $\sigma_{22}=\sigma_{33}=P$, in agreement with the
numerical findings.

In both of the above cases we have argued that the actual velocity
profiles are set by the sub-dominant dependence on the shear rate.
If true, we would expect strong fluctuations of the velocity around
the average profile, something observed in both systems described
above~\cite{Midi04,Silbert05b}.

\subsection{Modified Couette geometry}
In both of the linear geometries considered above, predicting the
shape of shear zones in terms of SFS is trivial since symmetry
guarantees that the SFS are parallel with the boundaries.
\note{Balancing the stresses is hence also trivial due to the
special form of the stress tensor in the SFS basis.} We now tackle
the modified Couette geometry. Compared with the examples considered
so far, this system has lost the symmetry in the $\v e_1$ direction
(along the SFS perpendicular to the shear; see
Fig.~\ref{fig:setups}). The remaining symmetry in the $\v e_2$
direction is either rotational, as in the case of the modified
Couette system (Fig.~\ref{fig:setups} c), or translational as in the
linear system (Fig.~\ref{fig:setups} d). Though the loss of symmetry
makes the treatment much more involved, it will lead to the
conclusion that a constant effective-friction coefficient is {\em
not} consistent with slow granular flows in general --- we will find
that the appropriate shape of SFS that describes the expected wide
shear zones do not occur when we have a constant effective friction
coefficient. We suggest that a dependence of the friction on the
local angles, as indicated in Eq.~(\ref{ang_dep}), is crucial to
understand such slow granular flows.

\subsubsection{Rotational symmetry along the shearing direction}
The system depicted in Fig.~\ref{fig:setups} c consists of a
cylindrical container filled with a granular material, and with a
split bottom plate. The inner part of the container is rotated at an
angular velocity $\Omega$, long enough for a steady state to be
reached. The key experimental finding regards the spread of a wide
shear band from the bottom slit up through the bulk to the surface.
Naturally most data was collected for the velocity profiles at the
top surface, as a function of the total height of the sample $H$.

It was found in~\cite{Fenistein03,Fenistein04} that the center
position of the shear zone, $R_{\rm c}$, and its width, $W$, satisfy
simple scaling relations as function of the layer height, $H$, the
radial position of the bottom slit, $R_{\rm s}$, and the grain
diameter, $a$. To good accuracy \be{eq:scale}
\begin{array}{rl}
 1-R_{\rm c}/R_{\rm s}=&(H/R_{\rm
 s})^{5/2},\\  W\propto &H^{2/3} a^{1/3}.
\end{array}
\ee Though the experiments naturally focused on the velocity
profiles at the top surface, there is also evidence that inside the
bulk, away from the surface, the width of the shear zone scales with
the height above the bottom $z$ as $W(z)\propto z^{\alpha}$, with
$\alpha$ somewhere between $0.2$ and
$0.4$.~\cite{Fenistein04,Fenistein05,Cheng05}.

In the natural cylindrical coordinate system, with the normalized
basis $\{\v e_r,\v e_\varphi, \v e_z\}$, we have $\v v= v\v
e_\varphi$, and thus \be{eq:17} (\vv D)_{{\rm
cyl}}=\frac{r}{2}\left(\begin{array}{ccc} 0& \partial_r \omega & 0\\
\partial_r \omega &0 &\partial_z \omega\\ 0& \partial_z \omega &
0\end{array}\right), \quad \omega=v/r.  \ee Due to the symmetry of
the problem, the surfaces of constant angular velocity are
identified as the SFS. By choosing the SFS basis
\begin{eqnarray}
\nonumber  \v e_1&=&\frac{1}{|\v \nabla \omega|}(\partial_z\omega \v e_r-\partial_r \omega \v e_z),\\
  \v e_{2}&=&\v e_\varphi,\\
\nonumber  \v e_{3}&=&\frac{1}{|\v \nabla \omega|}(\partial_r\omega
\v e_r+\partial_z\omega \v e_z),
\end{eqnarray}
we arrive at the right form of the strain-rate tensor~(\ref{eq:De}),
with $\dot\gamma=r|\v\nabla\omega|/2$. Due to the complicated
geometry we need to work with the full momentum-continuity
equations~(\ref{eq:qscons}) in order to proceed.
\begin{figure}
\includegraphics[width=.8\columnwidth]{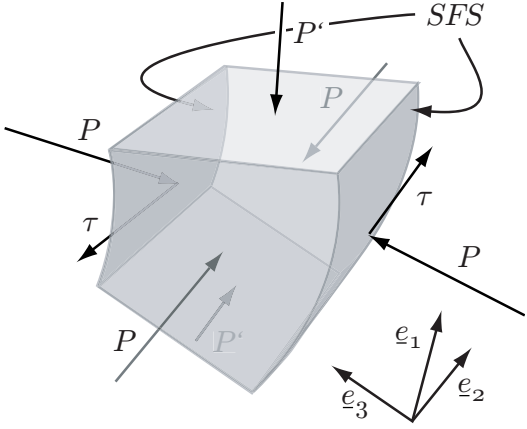}
\caption{\label{cuboid} Forces acting on a small element of material
sandwiched between two shear-free sheets (indicated). }
\end{figure}
We denote the derivative along the $\hat{e}_i$:th direction as
$\frac{d}{dx^i}:= \hat{e}_i \cdot \v{\nabla}$. In Fig.~\ref{cuboid}
we have sketched a local cuboidal element of material contained
between two SFS and illustrate the forces a stress tensor of the
form Eq.~(\ref{eq:stressr}) would give rise to.

We introduce the angle $\theta$ as the angle that $\v e_1$ makes with the
$z$-axis, $\kappa_1=\frac{\d \theta}{\d x^1}$ as the curvature of
the integral curves of $\v e_1$ (constant-$\omega$ curves), and
$\kappa_3=\frac{\d \theta}{\d x^3}$ as the curvature of the integral
curves of $\v e_3$. Taking care of the fact that the SFS basis forms
the normalized basis vectors of a curved coordinate system, and
using the machinery of tensor algebra, we can write
Eq.~(\ref{eq:qscons}) in the SFS basis as
\begin{eqnarray}
 \nonumber \frac{\d P'}{\d x^1}+(P-P')(\kappa_3-\sin\theta/r)&=&-\rho g\cos\theta,\\
  \frac{\d \tau}{\d x^3}+\left(\kappa_1-2\cos\theta/r\right)\tau&=&0,\label{eq:euler}\\
 \nonumber \frac{\d P}{\d x^3}+(P-P')\kappa_1&=&-\rho g\sin\theta.
\end{eqnarray}

\label{sec:curved} In the modified Couette setup depicted in
Fig.~\ref{fig:setups}c, we only need to specify $\theta$ to fix the
SFS basis. Hence the relations between stress components and angles
become of the form \be{eq:23} \tau=\mu(\theta) P, \quad
P'=\nu(\theta) P. \ee

The full equations~(\ref{eq:euler}) coupled to Eq.~(\ref{eq:23}) are
too complicated for a full analytical treatment. We therefore
will start from the simple assumption that the normal stress ratio
$\nu$ is equal to one (i.e., $P'\!=\!P$), and that $\mu$ is
constant, i.e., independent of $\theta$. Then we are left with the
equations
\begin{eqnarray}
 \nonumber   \frac{\d P}{\d x^1}&=&-\rho g\cos\theta,\\
  \frac{\d P}{\d x^3}+\left(\kappa_1-2\cos\theta/r\right)P&=&0,\\
 \nonumber   \frac{\d P}{\d x^3}&=&-\rho g \sin \theta.
\end{eqnarray}
The curvature $\kappa_3$ has dropped out, and the first and last
equations can be integrated to give a hydrostatic pressure profile
$P=\rho g(H-z)$. Upon substituting this into the second equation we
conclude that the curvature of the  constant-$\omega$ curves
satisfies,
\be{eq:24}
 \kappa_1=\frac{2\cos\theta}{r}+\frac{\sin\theta}{H-z}.
 \ee
To connect this formalism to the actual shapes of the SFS, let
$r(z)$ be the curves of constant $\omega$. Using that $\frac{dr}{dz}
= \tan{\theta}$ it follows that \be{eq:25}
\begin{array}{l}
  \sin\theta=\frac{r'(z)}{\sqrt{1+(r'(z))^2}}, \quad\cos\theta=\frac{1}{\sqrt{1+(r'(z))^2}},
\\\kappa_1=\frac{r''(z)}{(1+(r'(z))^2)^{3/2}}.
\end{array}
 \ee
 Hence using~(\ref{eq:24}) and (\ref{eq:25}) we see that the
curves of constant angular velocity must satisfy
\be{eq:alpha}
r''(z)=(1+(r'(z))^2)\left(\frac{2}{r}+\frac{r'(z)}{H-z}\right).
\ee
This is a second order differential equation which can be solved
numerically when supplemented with two boundary conditions. This
turns out to be exactly the same differential equation one arrives
at through minimizing the functional
\be{eq:26} f[r(\cdot)]=\int\limits_0^H\d z
(H-z) r^2(z)\sqrt{1+(r'(z))^2}.  \ee
As was shown by Unger and coworkers~\cite{Unger04}, this functional
can be seen to describe the torque needed to shear an ideal
cohesion-less Coulomb material under hydrostatic pressure which has
its infinitesimal shear zone at $r(z)$. Minimizing this torque, a
definite prediction for $r(z)$ was obtained which qualitatively
captures the shear zone location as measured experimentally. We
refer to~\cite{Unger04,Fenistein04,Fenistein05} for further
discussion.

\note{However, this approach cannot result in shear zones with a
width of the form seen in experiments, $W(z)\propto z^\alpha$, with
$.2<\alpha<.5$. To see this, one only has to consider the profiles
close to the bottom. We now assume that a general level curve has
the form~\footnote{Close to the bottom slit the system will no
longer be quasi-static due to large gradients in the velocity filed.
On the other hand, this region can be made arbitrarily small by
lowering the driving rate. Hence, a small-$z$ expansion should have
a finite region in which it accurately descries the behavior of the
SFS.}
\be{}
  r_1(z)=r_0(z)+A z^\alpha+h.o.t.,
\ee
where $r_0(z)$ is the center curve, and A some constant specifying
the specific level curve under consideration. Upon substituting this
into equation~(\ref{eq:alpha}), and considering the lowest order in
$z$, we conclude that $\alpha$ equals zero or one. This contradicts
the experimental findings.}

 From this we conclude that our assumptions that $\nu$ and $\mu$ are constant are not consistent
with the wide shear zones observed in the modified Couette geometry.

This curved geometry is, however, too complicated to study the
precise role of more general $\nu$ and $\mu$. We will therefore turn
our attention to the closely related linear split bottom shear cell,
which can be obtained by letting the slit radius diverge, and where
the rotational symmetry in the $\v e_2$ direction is replaced by a
simpler translational symmetry.

\subsubsection{Translation symmetry along the shearing direction}

\label{subsec:linear} 
\begin{figure}
\includegraphics[width=.8\columnwidth]{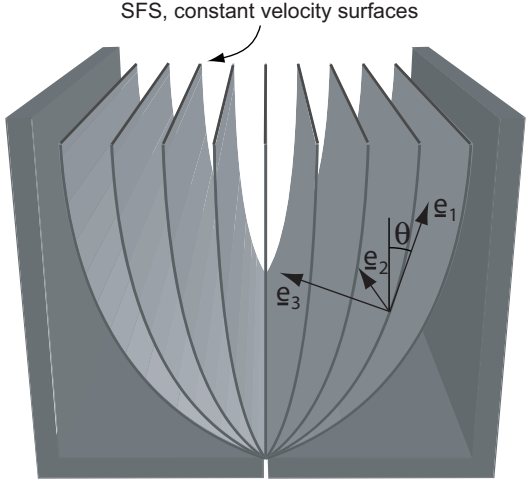}
\caption{\label{fig:linear} Schematic cross section of the linear
modified Couette geometry with the expected form of the SFS
indicated.}
\end{figure}

The scaling forms~(\ref{eq:scale}) relate the
shear zones width $W$, and location $R_{\rm c}$, to the particle
size $a$, height $H$, and radius of curvature of the slit $R_{\rm
s}$. Taking the limit $R_{\rm s} \rightarrow \infty$ enables us to
estimate what flow profiles can be expected in the linear setup
shown in Fig.~\ref{fig:setups} d, even though no experimental or
numerical data is available for such a system at
present~\footnote{Preliminary experiments, performed in Leiden, in a
conveyer belt geometry exhibit shear zones of finite width.}. The
width is independent of $R_{\rm s}$, while the shift between $R_{\rm
s}$ and $R_{\rm c}$ should vanish in this limit. We therefore expect
qualitatively the same widening of the shear zone as in the Couette
geometry (Fig.~\ref{fig:setups} c), with the shear zones center
remaining straight above the linear slit (consistent with the
reflection symmetry of such a linear geometry), as indicated in
Fig.~\ref{fig:linear}.

The equations for linear momentum conservation simplify to
\begin{eqnarray}
\nonumber  \frac{\d P'}{\d x^1}+(P-P')\kappa_3=-\rho g \cos\theta\\
  \frac{\d \tau}{\d x^3}+\kappa_1\tau=0\\
\nonumber  \frac{\d P}{\d x^3}+(P-P')\kappa_1=-\rho g \sin\theta.
\end{eqnarray}

Let us for the moment assume that $\mu$ is a constant, and test
whether this assumption is consistent with the flow profiles
sketched in Fig.~\ref{fig:linear}. For constant friction
coefficient, the last two equations above can be combined to give
\be{eq:27} P'\kappa_1=\rho g \sin \theta.  \ee For profiles that
bend upward when going through the bulk (see Fig.~\ref{fig:linear}),
$\kappa_1$ and $\theta$ are of opposite signs. Hence, to satisfy
Eq.~(\ref{eq:27}), $P'$ has to be negative, which is impossible in
cohesion-less granular materials.

\note{Thus we have two possible scenarios: If the effective friction
coefficient is constant, then the system can not support a wide
shear zone, and the shear must localize. In view of that the
cylindrical geometry exhibits a width of the shear zone that is
apparently independent of the position of the bottom slit, it seems
more likely though, that have upward bending profiles also in the
linear geometry. Hence, the effective friction coefficient must
decrease as we move away from the center, along the integral lines
of $\v e_3$.} This is a strong statement since it does not rely on
assuming any specific form for $\nu$, the ratio between the normal
pressures: Even with normal stress differences we can not get a
qualitatively correct description assuming the effective-friction
coefficient to be constant throughout the bulk.

\begin{figure}
\includegraphics[width=\columnwidth]{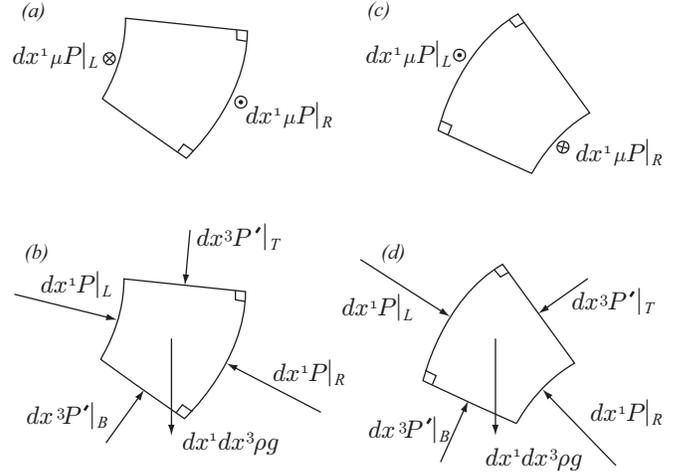}
\caption{\label{fig:pict} Illustration of the simple force balance
arguments which shows that a constant $\mu$ is incompatible with an
upward bending edge of the shear zone. For details, see text.
}
\end{figure}

The same conclusion can be reached by considering force balance on a
cuboid element of material contained between two SFS as illustrated
in Fig.~\ref{fig:pict} (a) and (b), and employing the special form
of the stress tensor in the SFS basis~(\ref{eq:stressr}). We start
from the fact that the total shear forces $dx^1 dx^2 \mu P$ acting
on the left and right edge of the cuboid need to balance. Now if
$\mu$ is a constant this implies that the normal forces on left and
right of the cuboid equal: $dx^1 dx^2 P|_L \!=\!dx^1 dx^2 P|_R$
(Fig.~\ref{fig:pict}a). The only additional forces acting on the
cuboid are gravity ($dx^1 dx^2 dx^3 \rho g$) and the normal forces
on top and bottom of the cuboid $dx^2 dx^3 P'|_{T,B}$
(Fig.~\ref{fig:pict}b). Due to the upward bending of the SFS, the
sum of these three terms clearly has a substantial component towards
the right --- hence it is impossible to balance forces in the $\v
e_3$ direction in this case. For ``outward'' bending SFS this
problems does not occur as illustrated in Fig.~\ref{fig:pict} (c)
and (d).

In our case, the only way to attain force balance is if the normal
force acting on the right face of the cuboid is larger than the
force acting from the left: $dx^1 dx^2 P'|_L \!<\!dx^1 dx^2 P'|_R$.
Since the shear forces still have to balance, this is only possible
when $\mu$ is not a constant --- in fact $\mu$ has to reach its {\em
maximum} along vertical SFS and then gradually decrease as we move
outward towards increasingly slanted SFS (along the $\v e_3$
direction).

The next step is thus to include a $\theta$-dependence in the
effective-friction coefficient and see if this is sufficient to be
able to obtain shear zones of finite width. For simplicity, we keep
 $\nu(\theta)=1$. As before, two of the momentum continuity
 equations can be solved and yield
 a hydrostatic pressure
profile, $P=\rho g(H-z)$. The third momentum continuity equation
becomes
\be{eq:full}
  \frac{\d \ln \mu(\theta)}{\d x^3}+\frac{\d \theta}{\d x^1}=\frac{ \sin\theta}{H-z}.
 \ee
To get an analytically tractable problem we now consider a region
close to the central level curve $r(z)=0$. Since odd powers of
$\theta$ can be excluded due to the $\theta\rightarrow -\theta$
symmetry, we assume that we can expand the friction coefficient as
\be{eq:28}
  \mu(\theta)=\mu_0\left(1-\frac{1}{2}q \theta^2\right)+\O(\theta^4).
 \ee
Sufficiently close to the central level curve $\theta$ is small, and
we can, to lowest order, rewrite the derivatives $\frac{\d}{\d x^1}$
and $\frac{\d}{\d x^3}$ as $\theta \partial_r + \partial_z$ and
$-\partial_r$ respectively. Hence the momentum conservation
equation~(\ref{eq:full}) can be rewritten, to lowest order, as
\be{eq:30}
  (1+q)\theta\frac{\partial \theta}{\partial r}+\frac{\partial \theta}{\partial z}=\frac{\theta}{H-z}+{\rm h.o.t.}.
 \ee
This differential equation can be solved by the method of
characteristics, resulting in
\be{eq:res}
  \theta(r,z)=-\frac{r/H}{(1+q)(1-z/H)\ln(1-z/H)}+{\rm h.o.t.}.
  \ee
Close to the bottom where $z/H\ll 1$, we expect the
constant-$\omega$ lines to satisfy $r(z)\propto z^{\alpha}$, with an
exponent $\alpha$ somewhere between $0.2$ and
$0.5$~\cite{Fenistein04}. In this limit, (\ref{eq:res}) can be
integrated using the fact that for any level curve we have
$r'(z)=\tan\theta(r(z),z)$. This results in
\be{eq:wide}
  r(z)\propto z^{1/(1+q)}, \quad z/H,\theta\ll 1. \ee The only
sensible profiles are achieved for $q>0$. This is in agreement with
the arguments sketched in Fig.~\ref{fig:pict}, indicating that
upward bending constant-$\omega$ lines are possible only if $\mu$
decreases with increasing $|\theta|$.  Hence, the highest
effective-friction coefficient is achieved when the direction of
gravity lies in the tangent plane of the shearing surfaces
\footnote{Close to the top surface, as well as close to the bottom
slit, the assumption of a small $\theta$ is no longer valid due to
the (slow) divergence in~(\ref{eq:res}), and the small $z$ behavior
of~(\ref{eq:wide}) respectively. On the other hand, we could never
hope to describe these regions with a quasi-static theory due to
that here $I$ (as given in Section~\ref{sec:dim}) becomes large, and
we leave the quasi-static part of the flow.}.

\begin{figure}
\includegraphics[width=\columnwidth]{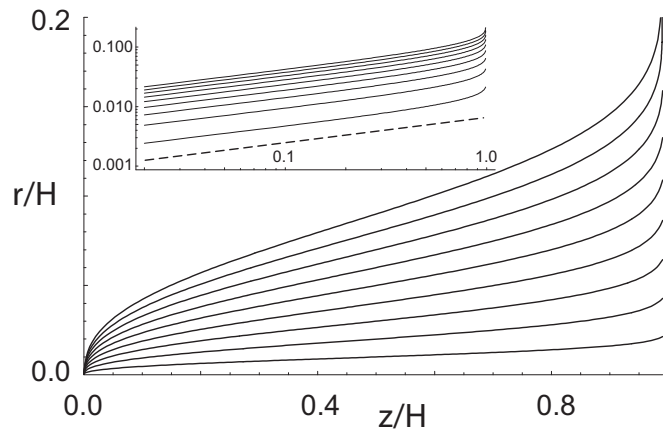}
\caption{\label{fig:data} Plots of the constant-$\omega$ lines as
calculated numerically in the small $\theta$
approximations~(\ref{eq:28}) for $q=1.5$, giving an exponent
$\alpha=1/(1+q)=0.4$ close to the bottom.  The inserted graph
depicts the same curves on a log-log scale, and with the $0.4$ power
indicated by the dashed line.}
\end{figure}

\section{Discussion and Conclusion}
\label{sec:discuss} To address the question of whether continuum
models can be made consistent with experimental and numerical
observations of wide shear zones in slow granular flows, we have
made a number of assumptions, which we will briefly recapitulate
here. Trivially, we assume that the flow profile is smooth on a
coarse grained scale. Furthermore we assume local action (see
Section~\ref{sec:qs}) and material objectivity (see
Section~\ref{sec:qs}), and focus on steady states for which the
shearing direction is a symmetry direction of the system --- this
washes out memory effects (see Section~\ref{sec:qs}). The considered
flows are sufficiently symmetric, and time-independent, so that they
can be described by time-independent Shear-Free Sheets. All these
assumptions appear rather inconspicuous.

But there are a number of less obvious assumptions which deserve
more attention, and for which a numerical test would be extremely
useful. The first of these is the absence of elastic shear stresses
in the flowing zone, due to rapid (on a macroscopic time scale)
relaxation of force fluctuations (see Section~\ref{sec:relax}).
Clearly, in a large system, far away from the shear zone, elastic
stresses should play a role, but here we only consider the actual
flowing region. It is an open question when and where such elastic
stresses start to play a role. Secondly, we assume that the packing
fraction is constant throughout the flowing region. Recent MRI
measurements of the packing density suggest an approximately
constant dilated region in the flowing zone \cite{Umbanhowar05}.
Nevertheless, far away from the shear zone the density is observed
to be different from this region. Finally we have excluded a
possible dependence of the effective-friction coefficient on the
pressure ratio, $\nu$, within the SFS (see Section~\ref{sec:dim}).
This assumption lacks a strong physical argument but is made to keep
the problem tractable, and should be an important issue to check
numerically

Using the assumptions recapitulated above, our method is based on
separating out those parameters of the strain-rate tensor that are
explicitly rate dependent. This enables us to build a explicitly
rate-independent theory, and we have shown that it is able to
predict some of the features of the stresses seen in numerical
simulations of the inclined plane geometry, as well as capturing the
widening of the shear zone in the modified Couette geometry.

Through the introduction of shear-free sheets we have also clarified
when a direct interpretation along the lines of solid friction can
be made, and further indicated how far such an analogy can be
stretched. Due to that the flow could be considered as consisting of
SFS, no special assumptions had to be made regarding the effect of a
variation in the principal-strain rate ratios throughout the sample.
It was further shown that in order to account for the expected shape
of the shear zones, the proportionality constants between the
different pressures (e.g. the effective-friction coefficient) must
retain a dependence on the local orientation of the flow (i.e. the
orientation of the principal-strain basis) relative to the local
body force --- the only other probable alternative is that the shear
zone is not wide. We speculate that the origin of such angle
dependence is due to the competition between the organizational
tendencies of the flow and the gravitational pull. The flow tends to
increase the number of grain contacts in compressional directions,
while decrease the number in expanding directions. At the same time,
the gravitational pull leads to an increased number of vertical,
opposed to horizontal connections (rattlers always fall down).
Unfortunately however, not enough is known about such angle
dependence of the contact network in order to confirm our
speculations. We suggest that this angle dependence as an important
issue for future research.

Due to the explicit rate independence of the approach, it can not
give the complete velocity profile. In order to determine the
complete profile one needs to include the sub-dominant rate
dependence in the stress tensor. This is straight forward for some
simple geometries and should be possible in general. Unfortunately
it turns out to be nontrivial even for the relatively simple
modified Couette geometry.

Nevertheless, the intriguing fact that the experimental shear
profiles in this geometry fitted an error function so well, provides
an important benchmark for understanding quasi-static flow. As we
have discussed in Section~\ref{subsec:linear}, a linear version of
this experiment may provide important additional information.

The present approach poses a set of well defined questions regarding
the packing fraction in the shear zone, the linear relationship
between pressures, the simple form of the stress tensor in the SFS
basis, and the dependence of the proportionality stress ratios $\mu$
and $\nu$, on the orientation of the shear planes with respect to
gravity. These are simple basic issues which are open to
investigation by numerical simulations, and possibly even by
experiments. Clarifying these issues appears crucial for further
development of a theory along these lines.

\section{Acknowledgments}
We gratefully acknowledge illuminating discussions with Ell\'ak
Somfai, Hans van Leeuwen, Wouter Ellenboek and Alexander Morozov.
M.D. acknowledges financial support from the physics foundation FOM
and PHYNECS, and MvH  acknowledges financial support from the
science foundation NWO through a VIDI grant.
\bibliography{shearbib}
\end{document}